# Skyrmions in Yang–Mills Theories with Massless Adjoint Quarks

R. Auzzi$^{(1)}$, S. Bolognesi$^{(2)}$ and M. Shifman$^{(2)}$

$^{(1)}$*Department of Physics, Swansea University,*
*Singleton Park, Swansea SA2 8PP, U.K.*

$^{(2)}$*William I. Fine Theoretical Physics Institute, University of Minnesota,*
*116 Church St. S.E., Minneapolis, MN 55455, USA*

**Abstract**

Dynamics of $SU(N_c)$ Yang–Mills theories with $N_f$ adjoint Weyl fermions is quite different from that of $SU(N_c)$ gauge theories with fundamental quarks. The symmetry breaking pattern is $SU(N_f) \to SO(N_f)$. The corresponding sigma model supports Skyrmions whose microscopic identification is not immediately clear. We address this issue as well as the issue of the Skyrmion stability. The case of $N_f = 2$ had been considered previously. Here we discuss $N_f \geq 3$. We discuss the coupling between the massless Goldstone bosons and massive composite fermions (with mass $O(N_c^0)$) from the standpoint of the low-energy chiral sigma model. We derive the Wess–Zumino–Novikov–Witten term and then determine Skyrmion statistics. We also determine their fermion number (mod 2) and observe an abnormal relation between the statistics and the fermion number. This explains the Skyrmion stability. In addition, we consider another microscopic theory – $SO(N_c)$ Yang–Mills with $N_f$ Weyl fermions in the vectorial representation – which has the same chiral symmetry breaking pattern and the same chiral Lagrangian. We discuss distinctive features of these two scenarios.

# 1  Introduction

In this paper we continue investigation of $SU(N_c)$ gauge theories with $N_f$ quarks in the adjoint representation. If the quarks are massless, the theory has the following pattern of chiral symmetry breaking ($\chi$SB) [1, 2, 3]:

$$SU(N_f) \times \mathbb{Z}_{2N_c N_f} \to SO(N_f) \times \mathbb{Z}_2, \qquad (1)$$

where the discrete factors are the remnants of the anomalous singlet axial U(1). The low-energy spectrum is described by a nonlinear sigma model with the target space given by the coset

$$\mathcal{M}_{N_f} = SU(N_f)/SO(N_f). \qquad (2)$$

This nonlinear sigma model is known to posses nontrivial topological solitons. In particular, it admits Skyrmions [4, 5, 6, 7] which can be classified by topologically nontrivial maps from the compactified three-dimensional space to the coset space. The relevant topological winding number is given by the third homotopy groups[1] which are collected in Table 1.

| $N_f$ | 2 | 3 | 4 | 5 |
|---|---|---|---|---|
| $\pi_3(\mathcal{M}_{N_f})$ | $\mathbb{Z}$ | $\mathbb{Z}_4$ | $\mathbb{Z}_2$ | $\mathbb{Z}_2$ |

Table 1: The third homotopy group for sigma models emerging in Yang–Mills theories with two, three, four and five adjoint flavors.

Unlike QCD, in the theories with adjoint quarks the Skyrmion mass scales as $N_c^2$ (here $N_c$ is the number of colors). Moreover, unlike QCD, where the relation between the Skyrmions and microscopic theory is well-established, this is not the case in the theories with adjoint quarks.

A natural question arises whether these Skyrmions which are topologically stable in the low-energy (macroscopic) description, are stable in the full (microscopic) theory. Since the low-energy Lagrangian is insensitive to the ultraviolet behavior of the microscopic theory, it is *apriori* possible that these Skyrmions might decay into the lowest-energy states of Yang–Mills with adjoint quarks through tunneling into these states.

---

[1] If $N_f \geq 6$ the theory is no longer asymptotically free. Thus, we limit ourselves to $N_f \leq 5$.

Thus, the issue of the Skyrmion stability in the full theory with adjoint quarks is nontrivial. The question was raised in [8, 9] in relation with the planar equivalence [10]. In theories with two index symmetric and antisymmetric quarks, this issue was addressed in Ref. [11]. For theories with adjoint matter, this question was already addressed in [12, 13]. The former paper was devoted to discussion of the (Hopf) Skyrmion stability in the simplest nontrivial example, $N_f = 2$. In the latter paper geometry of the $N_f = 3$ sigma model was analyzed. The purpose of the present paper is to complete the study of geometry of the coset space $\mathcal{M}_{N_f}$ for generic values of $N_f$, and reveal microscopic reasons for the Skyrmion stability in the theories with $N_f = 3, 4$ and 5. To this end we will need to consider peculiarities of interactions of color-singlet mesons with baryons with mass $O(N_c^0)$ present in the spectrum (the Skyrmion mass scales as $N_c^2$ in the theories under investigation).

More precisely, we will consider $N_f$ massless *Weyl* (or, which is the same, *Majorana*) fermions in the adjoint representation of the $\mathrm{SU}(N_c)$ gauge theory, to be denoted as $\lambda^a_{\alpha f}$ where $a$, $\alpha$ and $f$ are the color, Lorentz and flavor indices, respectively.[2] These Weyl fermions will be referred to as "quarks" or "adjoint quarks." To ensure flavor symmetry in the fundamental Lagrangian we take $N_f > 2$. At $N_f > 5$ asymptotic freedom is lost. The $N_f = 1$ theory is in fact $\mathcal{N} = 1$ super-Yang–Mills, it is gapped and has no Goldstone bosons in the physical spectrum.

Equation (1) can be elucidated as follows. In the vacuum the Lorentz-scalar bilinear $\lambda^a_{\alpha f}\lambda^{a\,\alpha}_g$ condenses,

$$\langle \lambda^a_{\alpha f}\,\lambda^{a\,\alpha}_g \rangle \sim \Lambda^3 \delta_{fg}\,. \tag{3}$$

The above order parameter stays intact under those transformations from $\mathrm{SU}(N_f)$ which are generated by matrices antisymmetric under transposition. In other words, the condensate (3) is invariant under transformations from the $\mathrm{SO}(N_f)$ subgroup. Thus, the low-energy pion Lagrangian is a nonlinear sigma model with the target space $\mathcal{M}_{N_f}$. Besides $\mathrm{SO}(N_f)$, the low-energy Lagrangian possesses a discrete symmetry: the $Z_2$ remnant of the axial flavor-singlet $U(1)$. We will call this the fermion number $F$ since it counts the number of fermions modulo 2. Thus, $(-1)^F$ is well-defined.

---

[2]Our notation is as follows. Three Pauli matrices acting on the space-time spinors are denoted by $\sigma_i$. We use $t_a$ to denote the generators of the flavor group $\mathrm{SU}(N_f)$ (these are the matrices we called $\tau_i$ in [12]). The matrices $t_a$ are $N_f \times N_f$ Hermitian matrices with vanishing trace, in total $N_f^2 - 1$ matrices. The antisymmetric ones form the closed subalgebra $\mathrm{SO}(N_f)$; their number is $N_f(N_f - 1)/2$. For example, for $N_f = 3$ they are the Gell-Mann matrices $t_2$, $t_5$, and $t_7$.

What is a crucial difference which makes the case of $N_f = 2$ easier than $N_f = 3$, 4 and 5? In this case a valid symmetry of the model surviving after $\chi$SB is SO(2) equivalent to U(1) [12]. One can classify all physical states with respect to the U(1) charge (it was referred to as $Q$ in [12]).

The low-energy effective Lagrangian describes dynamics of the lightest particles. The massless particles are the Nambu–Goldstone bosons $\pi$, which are in the 2-index symmetric traceless representation of SO($N_f$). In the fermionic sector the particle with the lowest mass is $\psi$, interpolated by the gauge invariant operator

$$\psi_{\beta\, f} = C \operatorname{Tr} \left( \lambda_f^\alpha F_{\alpha\beta} \right) \equiv C \operatorname{Tr} \left( \lambda_f^\alpha \sigma_{\alpha\beta}^{\mu\nu} F_{\mu\nu} \right), \tag{4}$$

were $F_{\alpha\beta}$ is the (anti)self-dual gluon field strength tensor (in the spinorial notation), and $C$ is a normalizing factor, $C \sim (N\Lambda^2)^{-1}$. The above composite fermions are in the vectorial representation of SO($N_f$).

The results of Ref. [12] can be summarized as follows. The Hilbert space accessible from the perturbative analysis is

$$\mathcal{H}^{(\text{hadronic})} = \mathcal{H}^{(+1,+1)} \oplus \mathcal{H}^{(-1,-1)} \tag{5}$$

containing the composite states with the even and odd U(1) charges, respectively. We have denoted the charges as $\{(-1)^Q, (-1)^F\}$. In particular, $\mathcal{H}^{(+1,+1)}$ contains the massless Nambu–Goldstone bosons $\pi^{\pm\pm}$ and, hence, there is no mass gap here. On the contrary, $\mathcal{H}^{(-1,-1)}$ has a mass gap $m$, the mass of the lightest composite fermion of the type (4). After analyzing the Skyrmion of the low-energy effective Lagrangian, we argued that an *extra* sector to which the Skyrmions belong is

$$\mathcal{H}^{(\text{exotic})} = \mathcal{H}^{(+1,-1)} \oplus \mathcal{H}^{(-1,+1)}. \tag{6}$$

Table 2 summarizes $Q$ and $F$ charges of various particles. From this Table it is seen that the Hopf Skyrmion stability is not a low-energy artifact. This is due to the fact that all conventional mesons and baryons with $m = O(N_c^0)$ in the theory at hand have

$$(-1)^Q \cdot (-1)^F = 1,$$

while for the Hopf Skyrmion

$$(-1)^Q \cdot (-1)^F = -1.$$

The reason behind this is that the fermion number has an anomalous contribution that couples directly to the topological current of the Skyrmion. This means that the Skyrmion acquires a fermion number through the so called Goldstone–Wilczek mechanism [14].

|  | $Q$, $N_f = 2$ | SO($N_f$), $N_f \geq 3$ | $F$ |
|---|---|---|---|
| $\psi$ | 1 | $\underline{N}_f$ | 1 |
| $\pi$ | 2 | 2-Tens Sym, Traceless | 0 |
| Skyrmion | 0 or 1 | 0 ——— | 1 0 |

Table 2: Global symmetry quantum numbers for nonexotic and exotic hadrons.

Now, what must be done to generalize this result to higher $N_f$? The generalization is not quite trivial. The residual symmetry which was U(1) in the $N_f = 2$ case is now replaced by SO($N_f$) with $N_f = 3, 4, 5$. Correspondingly, all particles from the physical spectrum must be classified according to representations of SO($N_f$), see Table 2. One can argue that for $N_f = 3, 4, 5$ the Goldstone–Wilczek mechanism provides the Skyrmion with an anomalous fermion number. Then we face a problem. For $N_f$ odd (i.e. $N_f = 3$ and 5), the quantum number assignments in Table 2 do *not* guarantee the stability of the Skyrmion. This is due to the existence of the antisymmetric tensor $\varepsilon^{i_1, i_2, \ldots, i_{N_f}}$ in SO($N_f$). Using this tensor we can assemble $N_f$ composite fermions $\psi$ in a combination invariant under the flavor group SO($N_f$), creating a baryonic final state. For $N_f$ odd this state would have the same quantum numbers as the Skyrmion and, thus, we could conclude that the Skyrmion, being an object with mass $\propto N_c^2$ would decay into $N_f$ composite fermions $\psi$ with mass $O(N_c^0)$ in the flavor singlet configuration.

An important role in the spin and statistics determination for solitons belongs to the Wess–Zumino–Novikov–Witten (WZNW) term [15]. To make the question more explicit it is instructive to briefly review the situation in conventional QCD with fundamental quarks [5, 6]. To begin with, let us consider two flavors, $N_f = 2$.

In this case the low-energy chiral Lagrangian does not admit the WZNW term. It does support Skyrmions, however. After quantization the Skyrmion quantum numbers $(I, J)$ form the following tower of possible values: $(0, 0)$, $(1/2, 1/2)$, $(1, 1)$, $(3/2, 3/2)$, etc. Here $I$, $J$ stand for isospin and spin, respectively. In the absence of the WZNW term Skyrmions can be treated as both, bosons and fermions. This is due to the fact that we may or may not add an extra sign in the field configurations belonging to nontrivial maps of $\pi_4(SU(2))$ [16].

At $N_f \geq 3$ the choice of the Skyrmion statistics (i.e. boson vs. fermion) becomes unambiguous. The reason is well-known: at $N_f \geq 3$ it is possible (in fact, necessary) to introduce the WZNW term in the effective Lagrangian [5, 6]. This term, which is absolutely essential in the anomaly matching between the ultraviolet (microscopic) and infrared (macroscopic) degrees of freedom, is responsible for the spin/statistics assignment for Skyrmions.

A similar situation takes place in adjoint QCD. With two flavors the WZNW term does not exist since $\pi_4(\text{SU}(2)/\text{U}(1)) = Z_2$. Quantization [17] gives us two possible tower of states: bosons with the U(1) charge and spin $(0, 0)$, $(2, 1)$, and so on, and fermions with the U(1) charge and spin $(1, 1/2)$, $(3, 3/2)$, and so on (see Table 2). In the effective low-energy theory it is impossible to decide in which of the two towers the Skyrmion lies. Only considering higher $N_f$ can we answer this question. The answer will play a crucial role in the explanation of the Skyrmion stability.

As well known [6], the SO($N_c$) gauge theory with $N_f$ Weyl fermions in the vectorial representation has the same as in Eq. (1) pattern of the global symmetry breaking, and is also described by a nonlinear sigma model with the target space $\mathcal{M}_{N_f}$. Witten proposed [6] that the Skyrmions of this theory must be identified with objects obtained by contracting the SO($N_c$) antisymmetric tensor $\varepsilon_{\alpha_1...\alpha_{N_c}}$ with the color indices of the vectorial quarks and/or the gluon field strength tensor (see Eqs. (63) and (64) below). These objects are stable due to the quotient symmetry $\mathbb{Z}_2 = \text{O}(N_c)/\text{SO}(N_c)$, which acts as a global symmetry group. Our results for the Skyrmion statistics in adjoint QCD can be applied to the SO($N_c$) gauge theory too. They give further evidence in favor of Witten's identification.

In this paper we investigate the reasons explaining the Skyrmion stability in adjoint QCD with $N_f \geq 3$ from the standpoint of the microscopic theory. We find that the Skyrmions are stable since they are the only particles with an odd relation between statistics and fermion number. Namely, Skyrmions can be bosons with with fermion number one or fermions with the vanishing fermion number. Therefore, Skyrmions cannot decay to any final state consisting of "normal" or "perturbative" particles: pions and other similar mesons or baryons of the type (4).

The paper is organized as follows. In Sect. 2 we describe in detail the low-energy effective action, parametrization of the manifold $\mathcal{M}_{N_f}$, and introduce a coupling to baryons $\psi$ of the type (4). Section 3 is devoted to determination of the WZNW term and calculation of its coefficient through the anomaly matching. Section 4 describes the effect of the WZNW term on the spin/statistics and fermion number of the Skyrmion. In Sect. 5 we describe the relevance of our results for another theory with the same global symmetry breaking pattern, SO($N_c$) QCD with $N_f$ Weyl fermions in the vectorial representation. In Sect. 6 we discuss an anomalous term responsible for the shift of the Skyrmion fermion number, which, in turn, guarantees its stability. In Sect. 7 we briefly touch upon the issue of the flux tubes supported by the chiral sigma model with the target space (2). Section 8 summarizes main conclusions of the paper. In Appendix we present technical details of the calculation needed in Sect. 3.

## 2 Low-Energy Effective Action

In this section we describe in detail the low-energy effective action and geometry of the coset manifold (2).

### 2.1 The coset space

We use the Cartan embedding to parametrize the coset (2). This parametrization is very useful, in particular, because it makes explicit the symmetries of the manifold. Then, coupling pions to baryons (4) becomes a straightforward task through the Cartan embedding.

The general element of the quotient $\mathcal{M}_{N_f} = \mathrm{SU}(N_f)/\mathrm{SO}(N_f)$ can be written in a compact form as $U \cdot \mathrm{SO}(N_f)$, where $U$ is an $\mathrm{SU}(N_f)$ matrix (different $U$ in $\mathrm{SU}(N_f)$ correspond to the same $\mathcal{M}_{N_f}$ element, modulo a product with an arbitrary $\mathrm{SO}(N_f)$ element). The map

$$U \cdot \mathrm{SO}(N_f) \to W = U \cdot U^t, \tag{7}$$

where the superscript $t$ denotes transposition, is well-defined on the quotient because for the $\mathrm{SO}(N_f)$ matrices the inverse is equal to the transposed matrix. Equation (7) presents a one-to-one map between $\mathcal{M}_{N_f}$ and the submanifold of the matrices of $\mathrm{SU}(N_f)$ which are both unitary and symmetric. In the mathematical literature it is called the Cartan embedding (see Ref. [18] for a review).

The quotient (2) can be parameterized using the matrix exponential of the $SU(N_f)$ generators which do *not* belong to the unbroken subgroup $SO(N_f)$. These generators are given by $N_f \times N_f$ *symmetric* Hermitian matrices with vanishing trace. These matrices necessarily have all real entries; they can be always diagonalized by virtue of an $SO(N_f)$ change of basis. Thus, a convenient way of parameterization of the equivalence classes of $\mathcal{M}_{N_f}$ is as follows: one must choose an $SU(N_f)$ representative

$$U = \exp(i\, V \cdot A \cdot V^\dagger), \tag{8}$$

where $A$ is a Hermitian traceless diagonal matrix and $V$ is an $SO(N_f)$ element, which can be parameterized by $N_f$-dimensional Euler angles. Then we can construct the matrix $W$,

$$W = U \cdot U^t = \exp\left(2i\, V \cdot A \cdot V^\dagger\right). \tag{9}$$

The Lagrangian of the Skyrme model with the target space $\mathcal{M}_{N_f}$ can be computed by evaluating the Lagrangian of the $SU(N_f)$ Skyrme model on the symmetric unitary matrix $W$,

$$\begin{aligned}
\mathcal{L} &= \frac{F_\pi^2}{4}\mathcal{L}_2 + \frac{1}{e^2}\mathcal{L}_4 \\
&\equiv \frac{F_\pi^2}{4}\mathrm{Tr}\left(\partial_\mu W \partial^\mu W^\dagger\right) + \frac{1}{e^2}\mathrm{Tr}\left[(\partial_\mu W)W^\dagger, (\partial_\nu W)W^\dagger\right]^2.
\end{aligned} \tag{10}$$

For instance, for $N_f = 2$ a straightforward parameterization of $A$ and $V$ is

$$A = \begin{pmatrix} +\theta/2 & 0 \\ 0 & -\theta/2 \end{pmatrix}, \quad V = \begin{pmatrix} \cos\alpha/2 & -\sin\alpha/2 \\ \sin\alpha/2 & \cos\alpha/2 \end{pmatrix}. \tag{11}$$

Now we can compute the Cartan embedding,

$$W = U \cdot U^t = \begin{pmatrix} \cos\theta + i\sin\theta\cos\alpha & i\sin\theta\sin\alpha \\ i\sin\theta\sin\alpha & \cos\theta - i\sin\theta\cos\alpha \end{pmatrix}. \tag{12}$$

It is not difficult to check that Eq. (10) gives the standard $S^2$ Skyrme model.

For $N_f = 3$ we can use the parameterization suggested in Ref. [13],

$$A = \frac{1}{2}\begin{pmatrix} \eta/\sqrt{3} + \theta & 0 & 0 \\ 0 & \eta/\sqrt{3} - \theta & 0 \\ 0 & 0 & -2\eta/\sqrt{3} \end{pmatrix}, \tag{13}$$

$$V = \begin{pmatrix} \cos\frac{\alpha}{2}\cos\frac{\gamma}{2} - \cos\frac{\beta}{2}\sin\frac{\alpha}{2}\sin\frac{\gamma}{2} & -\sin\frac{\alpha}{2}\cos\frac{\gamma}{2} - \cos\frac{\alpha}{2}\cos\frac{\beta}{2}\sin\frac{\gamma}{2} & \sin\frac{\beta}{2}\sin\frac{\gamma}{2} \\ \cos\frac{\alpha}{2}\sin\frac{\gamma}{2} + \cos\frac{\beta}{2}\sin\frac{\alpha}{2}\cos\frac{\gamma}{2} & -\sin\frac{\alpha}{2}\sin\frac{\gamma}{2} + \cos\frac{\alpha}{2}\cos\frac{\beta}{2}\cos\frac{\gamma}{2} & -\cos\frac{\gamma}{2}\sin\frac{\beta}{2} \\ \sin\frac{\alpha}{2}\sin\frac{\beta}{2} & \cos\frac{\alpha}{2}\sin\frac{\beta}{2} & \cos\frac{\beta}{2} \end{pmatrix}. \tag{14}$$

The angle variation range for $\theta$ and $\eta$ is

$$0 \leq \theta \leq, \pi \quad -\frac{\theta}{\sqrt{3}} \leq \eta \leq \frac{\theta}{\sqrt{3}}. \tag{15}$$

The range of variation for the Euler parameters is

$$0 \leq \alpha \leq 2\pi, \quad 0 \leq \beta \leq 2\pi, \quad 0 \leq \gamma \leq 2\pi. \tag{16}$$

It is straightforward to explicitly compute the metric of the quotient manifold. To this end we need just to evaluate Eq. (10) on $W$,

$$\mathcal{L}_2 = \frac{1}{4}\bigg[2(\partial_\mu\theta)^2 + 2(\partial_\mu\eta)^2 + 2\sin^2\theta(\partial_\mu\alpha)^2$$

$$+(1 - \cos\sqrt{3}\eta\cos\theta - \cos\alpha\sin\sqrt{3}\eta\sin\theta)(\partial_\mu\beta)^2$$

$$+\frac{1}{2}(\partial_\mu\gamma)^2\bigg[2 - (1 + \cos\beta)\cos^2\theta - 2\cos\sqrt{3}\eta\cos\theta\sin^2\frac{\beta}{2}$$

$$+2\cos\alpha\sin^2\frac{\beta}{2}\sin\sqrt{3}\eta\sin\theta + \sin^2\theta + \cos\beta\sin^2\theta\bigg]$$

$$+\left(4\cos\frac{\beta}{2}\sin^2\theta\right)(\partial_\mu\alpha)(\partial_\mu\gamma) - \left(2\sin\alpha\sin\frac{\beta}{2}\sin\sqrt{3}\eta\sin\theta\right)(\partial_\mu\beta)(\partial_\mu\gamma)\bigg]. \tag{17}$$

This is an independent check of the result found in Ref. [13] using another method. It is straightforward to generalize this parametrization to higher $N_f$. Then $A$ is just a generic $N_f \times N_f$ diagonal traceless real matrix and $V$ is an SO($N_f$) element parameterized by $N_f$-dimensional Euler angles.

The formulation in terms of $A$ and $V$ is suitable for description of the global structure of the manifold (2), but it is singular near $A = 0$. For many purposes it is more convenient to use the physical pion fields $\pi_k$. For $N_f = 3$ this reduces to

$$W = \exp[2i\, V \cdot A \cdot V^\dagger] = \exp\left[\frac{2i}{F_\pi}\begin{pmatrix} \pi_1 + \frac{\pi_3}{\sqrt{3}} & \pi_2 & \pi_4 \\ \pi_2 & -\pi_1 + \frac{\pi_3}{\sqrt{3}} & \pi_5 \\ \pi_4 & \pi_5 & -2\frac{\pi_3}{\sqrt{3}} \end{pmatrix}\right]. \quad (18)$$

## 2.2 Equivalent formulation: gauged SO($N_f$)

We introduce now a different, but equivalent, formulation that will be particularly useful for the purposes addressed in Sect. 6. In the $N_f = 2$ case there are two ways to parametrize the target space $S^2$. One can use a vector $\vec{n}$ subject to the constraint $|\vec{n}| = 1$. This is the so-called O(3) formulation. Another approach, which goes under the name of the gauged formulation of the $CP^1$ sigma model, is to use a complex doublet $z_i$ subject to the constraint $z_i^* z_i = 1$. This leaves us with an $S^3$ sphere. We have to further reduce it by gauging the phase rotation $z_i \to e^{i\theta} z_i$. This Hopf fibration leaves us exactly with the sphere $S^2$. The map between the two formulations is

$$\vec{n} = z_i^* \vec{\tau} z_i\,.$$

The derivatives acting on the doublet $z_i$ are the covariant derivatives

$$D_\mu = \partial_\mu - i A_\mu$$

where (see e.g. [19] for a review)

$$A_\mu = -\frac{i}{2}\left[z_i^*(\partial_\mu z_i) - (\partial_\mu z_i^*) z_i\right]. \quad (19)$$

To avoid confusion in the case of higher $N_f$ we will refer to the latter formulations as the $z$ formulation.

The $z_i$ formulation has a valuable advantage. It is possible to express the Hopf charge (the charge of $\pi_3(S^2) = Z$) as a local function of the gauge field $A$,

$$s = \frac{1}{4\pi^2} \int d^3x \epsilon^{\mu\nu\rho} A_\mu \partial_\nu A_\rho. \tag{20}$$

An equivalent local expression in terms of the $\vec{n}$ field is impossible [20].

Generalization to higher $N_f$ *is not* achieved by extending the doublet to a complex $N_f$-plet. For $N_f = 2$ this strategy works because SU(2) is equivalent to the sphere $S^3$. In order to generalize to higher $N_f$ we need to start with an SU($N_f$) sigma model and then gauge an SO($N_f$) subgroup. Let us consider the exact sequence

$$\ldots \to \pi_3\left(\mathrm{SO}(k)\right) \to \pi_3\left(\mathrm{SU}(k)\right) \to \pi_3\left(\mathrm{SU}(k)/\mathrm{SO}(k)\right) \to \pi_2\left(\mathrm{SO}(k)\right) \to \ldots$$

For every $k$ we have $\pi_2\left(\mathrm{SO}(k)\right) = 0$. Therefore, every non-zero element of $\pi_3\left(\mathrm{SU}(k)/\mathrm{SO}(k)\right)$ can be lifted to a non-zero element of $\pi_3\left(\mathrm{SU}(k)\right)$ (for $N_f > 2$ this lifting is not unique, as we will discuss below). Then we can calculate the $S^3$ winding number of the lifted 3-cycle, using the SU($N_f$) result,

$$s = -\frac{i}{24\pi^2} \int_{S^3} \mathrm{Tr}\,(U^\dagger dU)^3. \tag{21}$$

For $N_f = 2$, this gives us the Hopf winding number. For $N_f = 3$ this number is defined modulo 4, because

$$\pi_3\left(\mathrm{SU}(3)/\mathrm{SO}(3)\right) = \mathbb{Z}_4.$$

For $N_f > 3$ this number is defined modulo 2, because

$$\pi_3\left(\mathrm{SU}(k)/\mathrm{SO}(k)\right) = \mathbb{Z}_2, \qquad k > 3.$$

This is due to the fact that there are different ways of performing lifting from SU($k$)/SO($k$) to SU($k$). They differ from each other by an element of $\pi_3\left(\mathrm{SO}(k)\right)$. These are indeed the topological configurations relevant for SO($k$) instantons. They will enter when extending the space from SU($k$)/SO($k$) to SU($N_f$) with the subgroup SO($N_f$) gauged.

An intuitive picture is shown in Fig. 1. Let us consider for simplicity $N_f \geq 4$. In this case $\pi_3(\mathcal{M}_{N_f}) = Z_2$. In the sigma model with the target space $\mathcal{M}_{N_f}$ topological vortices are present. They are associated with the second homotopy group $\pi_2$, which is $\mathbb{Z}$ for $N_f = 2$ and $\mathbb{Z}_2$ for $N_f \geq 3$. As discussed in Ref. [21], the Hopf Skyrmion can be interpreted in term of an adiabatic twist of this vortex. If we have a double adiabatic twist of the vortex, this configuration can be continuously transformed in the identity, if we leave the subspace $S^2$ and move into $\pi_3(\mathcal{M}_{N_f})$. It seems that we have a problem. The integral of the Hopf charge gets continuously transformed from 2, when evaluated on the double twisted vortex, to 0 when evaluated on the no-twist configuration. How is it possible? The answer, as explained in Figure 1, is that an instanton of $SO(N_f)$ is generated when we perform the "unwinding" of the Hopf Skyrmion of topological charge 2. (Remember that in order to define a Hopf charge we need to extend the space to $SU(N_f)$ with a subgroup $SO(N_f)$ gauged.) For $N_f = 3$ the situation is very similar but now the minimal instanton is doubled (see [22]). For this reason the twists can disappear only in multiples of 4. Note that for $N_f = 2$ the SO(2) theory is Abelian and thus there are no instantons; the number of twists is strictly conserved.

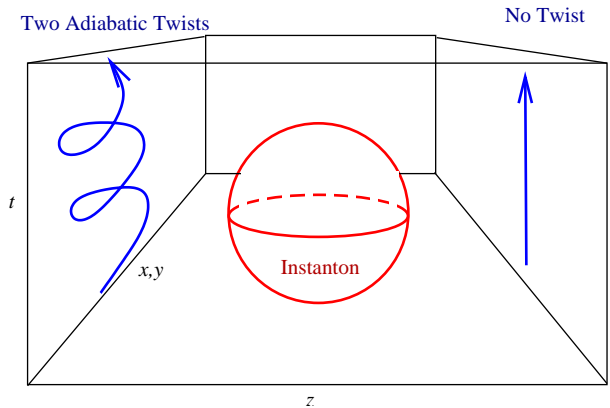

Figure 1: The cube represents a four-dimensional space. The vertical coordinate is time $t$. We perform an adiabatic twist in $t$. The depth corresponds to the plane $x,y$ where the vortex lies. The horizontal coordinate corresponds to $z$. On the left, in the space $(x, y, t, z = -\infty)$, we have a double-twisted vortex. On the right, at $(x, y, t, z = +\infty)$, we have a vortex with no twist. In the middle an instanton of $SO(N_f)$ is generated. This explains why the integral over the Chern–Simons term gets continuously changed from 2 to 0 for $N_f > 3$ and from 4 to 0 for $N_f = 3$.

It is possible to present the topological winding number as an $SU(N_f)$

Chern–Simons current. Let us introduce

$$\mathcal{A}_\mu = iU^\dagger \partial_\mu U \,. \tag{22}$$

Then

$$s = \frac{1}{8\pi^2} \int d^3x K^0, \quad K^\mu = \epsilon^{\mu\nu\rho\sigma} \operatorname{Tr}\left(\mathcal{A}_\nu \partial_\rho \mathcal{A}_\sigma - \frac{2}{3} i\, \mathcal{A}_\nu \mathcal{A}_\rho \mathcal{A}_\sigma\right). \tag{23}$$

As previously discussed, $s$ is defined modulo 4 for $N_f = 3$ and modulo 2 for $N_f > 3$, due to arbitrariness in the choice of $U$.

### 2.3  Fermion interaction

For $N_f = 2$ the coupling of pions to composite fermions (4) was considered in Ref. [12],

$$-\frac{g}{2}\left\{\psi^{\alpha f} \vec{n} \cdot (\vec{\tau})^g_f\, \psi_{\alpha g} + \text{h.c.}\right\}. \tag{24}$$

In order to generalize this to the case $N_f > 2$ one must express the coset (2) in a way that makes "evident" the action of the $\mathrm{SU}(N_f)$ symmetry on the coset. In the case of $\mathrm{SU}(2)/\mathrm{U}(1)$ it was easy since using the $n$ representation, with the unit vector $\vec{n}$, makes evident how it transforms under $\mathrm{SU}(2)$ rotations. However, the $N_f = 2$ case can be somewhat misleading for generalization to higher $N_f$.

$\mathrm{SU}(2)$ can be represented as the sphere $S^3$ in the four-dimensional vector space generated by the identity and the Pauli matrices $\sigma_i$. Intersecting this sphere with the hyperplane generated by the Pauli matrices we get an $S^2$ that is in one-to-one correspondence with the coset space $\mathrm{SU}(2)/\mathrm{U}(1)$. Moreover, this intersection tells us exactly how the $\mathrm{SU}(2)$ symmetry acts on the coset, it is the space $\{\vec{n}\}$ of unit vectors. Another possible way is to intersect the space with the hyperplane of the symmetric matrices generated by $1$, $\sigma_1$, $\sigma_3$. This is again a sphere $S^2$ and is again in one-to-one correspondence with the coset manifold. There is no contradiction with the symmetry properties since for $\mathrm{SU}(2)$ the adjoint representation is equivalent to the two-index symmetric and traceless representation. This is a consequence of equivalence between the fundamental and the antifundamental representations in $\mathrm{SU}(2)$.

To generalize this construction to higher $N_f$ we have to intersect the group $\mathrm{SU}(N_f)$ by the hyperplane generated by symmetric matrices. The space we get is in one-to-one correspondence with the coset $\mathcal{M}_{N_f}$ and is an explicit realization of its symmetric properties under the action of the SU($N_f$) group. We have thus a two-index symmetric matrix that can be saturated by the fermion bilinear $\psi^{a\alpha}\psi^{b\beta}\epsilon_{\alpha\beta}$. Once we orient the element of the coset manifold in one particular direction and expand around it, we get a Lagrangian with a manifest SO($N_f$) symmetry containing massless pions in the zero-trace two-index symmetric representation and one Weyl fermion in the fundamental representation $\underline{N_f}$. In addition to the kinetic term, the fermion has also an SO($N_f$) invariant mass term proportional to $\psi^{a\alpha}\psi^{b\beta}\delta_{ab}\epsilon_{\alpha\beta}$. Note that convolution of the indices $a,b$ with $\delta_{ab}$ is an operation that breaks the SU($N_f$) symmetry but is perfectly allowed in SO($N_f$). There are also interaction terms and higher derivative terms, all respecting the SO($N_f$) symmetry, which we do not need to know exactly for our purposes.

Let us consider an SU($N_f$) representative $U$ of a quotient class in $\mathcal{M}_{N_f}$. The SU($N_f$) symmetry group acts on $U$ as $W \to R \cdot U$. The action on the Cartan embedding image ($W = U \cdot U^t$) is

$$W \to R \cdot W \cdot R^t\,. \tag{25}$$

Due to this property, we can write down the fermion coupling for arbitrary $N_f$ as

$$-\frac{g}{2}\left\{W^{fg}\psi_{\alpha f}\psi_g^\alpha + \text{H.c.}\right\}. \tag{26}$$

To the lowest order, the effective Lagrangian which includes both pions and the fermions $\psi_{\alpha a}$ is

$$\mathcal{L} = \frac{F_\pi^2}{4}\text{Tr}\left(\partial_\mu W \partial^\mu W^\dagger\right) + \bar{\psi}_{f\dot\alpha}i\partial^{\dot\alpha\alpha}\psi_{f\alpha} - \frac{g}{2}\left\{W^{fg}\psi_{\alpha f}\psi_g^\alpha + \text{H.c.}\right\}. \tag{27}$$

If we expand around the vacuum where $W$ is given by the identity matrix, the fermionic part of the Lagrangian is given by

$$\mathcal{L}_{\text{ferm}} = \bar{\psi}_{f\dot\alpha}i\partial^{\dot\alpha\alpha}\psi_{f\alpha} - g\left\{\psi_f^\alpha\psi_{\alpha f} + \text{H.c.}\right\}. \tag{28}$$

Of course, there are interactions between these fermions and the Goldstone bosons. For example, for $N_f = 3$, in the first nontrivial order they are

$$\frac{2i}{F_\pi}\begin{pmatrix}\psi_1^\alpha & \psi_2^\alpha & \psi_3^\alpha\end{pmatrix}\begin{pmatrix}\pi_1 + \frac{\pi_3}{\sqrt{3}} & \pi_2 & \pi_4 \\ \pi_2 & -\pi_1 + \frac{\pi_3}{\sqrt{3}} & \pi_5 \\ \pi_4 & \pi_5 & -2\frac{\pi_3}{\sqrt{3}}\end{pmatrix}\begin{pmatrix}\psi_{\alpha 1} \\ \psi_{\alpha 2} \\ \psi_{\alpha 3}\end{pmatrix}. \tag{29}$$

Following Ref. [12] it is convenient to transfer this interaction from the potential to the kinetic term. Let us introduce the SU($N_f$) matrix $\tilde{U}$ such that

$$\tilde{U}^t(x) \cdot W(x) \cdot \tilde{U}(x) = 1 \,. \tag{30}$$

The matrix $\tilde{U}$ can be chosen as

$$\tilde{U}(x) = \exp\left(i\,\nu_k(x)\,\lambda_k\right) \,, \tag{31}$$

where $\lambda_k$ are the broken symmetric generators of SU($N_f$). Moreover, let us introduce a fermion variable $\chi$ such that

$$\psi = \tilde{U}\,\chi \,, \qquad \bar{\psi} = \bar{\chi}\,\tilde{U}^\dagger \,. \tag{32}$$

With these variables, the fermion part of Eq. (27) takes the form

$$\mathcal{L}_{\text{ferm}} = \bar{\chi}_{f\dot{\alpha}}\left(i\partial^{\dot{\alpha}\alpha} + \tilde{\mathcal{A}}^{\dot{\alpha}\alpha}\right)\chi_{f\alpha} - g\left\{\chi_f^\alpha \chi_{\alpha f} + \text{H.c.}\right\} , \tag{33}$$

where

$$\tilde{\mathcal{A}}_\mu = i\tilde{U}^\dagger \partial_\mu \tilde{U} \,. \tag{34}$$

In the first nontrivial order we have

$$\tilde{\mathcal{A}}_\mu \approx i\left(\partial_\mu \nu_k\right)\lambda_k + \frac{1}{2}\left[\nu_k \lambda_k, (\partial_\mu \nu_j)\lambda_j\right] \,. \tag{35}$$

It is easy to check that, if the current in Eq. (23) is calculated on $\tilde{\mathcal{A}}_\mu$, the SO($N_f$) components (and only SO($N_f$) components) of $\tilde{\mathcal{A}}_\mu$ give a nonvanishing contribution to the integral. This is due to the fact that the anticommutator of two $\lambda_k$ matrices is always an SO($N_f$) generator.

## 3 WZNW Term and Anomaly Matching

### 3.1 The Wess–Zumino–Novikov–Witten term

We can write the WZNW term for the $\mathcal{M}_{N_f}$ sigma model ($N_f \geq 3$) by virtue of evaluating the SU($N_f$) Wess–Zumino–Novikov–Witten term on the symmetric unitary matrices $W$ introduced in Eqs. (8) and (9). Namely,

$$\begin{aligned}
\Gamma &= -\frac{i}{240\pi^2}\int_{B_5} d\Sigma^{\mu\nu\rho\sigma\lambda}\text{Tr}\left[(W^\dagger \partial_\mu W)\cdot (W^\dagger \partial_\nu W)\right. \\
&\quad \cdot \left. (W^\dagger \partial_\rho W)\cdot (W^\dagger \partial_\sigma W)\cdot (W^\dagger \partial_\lambda W)\right] . \tag{36}
\end{aligned}$$

In order to compute the WZNW term for the $\mathcal{M}_{N_f}$ sigma model, we need to take the result for $\mathrm{SU}(N_f)$ and restrict it to the submanifold of the unitary symmetric matrices.

There is a subtle difference regarding the possible coefficients allowed for $\Gamma$ in the action. In the Lagrangian of the $\mathrm{SU}(N_f)$ sigma model, relevant for QCD Skyrmions, the WZNW term must have just integer coefficient $k$,

$$\mathcal{L} = \mathcal{L}_2 + k\,\Gamma + \text{Higher order terms}\,. \tag{37}$$

This is due to the fact that the integral of this term on an arbitrary $S^5$ submanifold of $\mathrm{SU}(N_f)$ must be an integer multiple of $2\pi$. In the $\mathcal{M}_{N_f}$ sigma model relevant for adjoint QCD we need to use the same normalization prescription. The main difference is that if we integrate $\Gamma$ on the minimal $S^5$ which we can build inside the $\mathrm{SU}(N_f)$ subspace of the symmetric Hermitian matrices, the result will be $4\pi$ rather than $2\pi$, as we get for the generator of $\pi_5\left(\mathrm{SU}(N_f)\right)$ (for an explicit calculation see Appendix). Therefore, if we restrict ourself to this subspace it is consistent to consider also half-integer values of $k$.

For $N_f = 4$ there is in principle a topological obstruction, since

$$\pi_4(\mathrm{SU}(4)/\mathrm{SO}(4)) = \mathbb{Z}\,. \tag{38}$$

A nonvanishing $\pi_4$ is not a problem for writing the WZNW term as it was for $N_f = 2$. It is possible to solve this problem embedding the $\mathrm{SU}(4)/\mathrm{SO}(4)$ space in $\mathrm{SU}(5)/\mathrm{SO}(5)$. If we did the same for $N_f = 2$, it would not work because the WZNW term is trivially zero on the manifold with dimension less than 5.

## 3.2 Gauged WZNW

The Cartan embedding map

$$U \cdot \mathrm{SO}(N_f) \to W = U \cdot U^t$$

gives us a realization of $\mathcal{M}_{N_f}$ as the submanifold of the symmetric elements of $\mathrm{SU}(N_f)$. Under an $\mathrm{SU}(N_f)$ element $R$

$$U \cdot \mathrm{SO}(N_f) \to R \cdot U \cdot \mathrm{SO}(N_f)\,, \qquad W \to R \cdot W \cdot R^t\,.$$

Therefore, if $R$ is an element of the unbroken $\mathrm{SO}(N_f)$ subgroup, it is mapped to a vectorial element of the $\mathrm{SU}(N_f)_R \times \mathrm{SU}(N_f)_L$ symmetry. If we gauge an element of the unbroken group in the $\mathcal{M}_{N_f}$ chiral Lagrangian, we can use the result for the corresponding $SO(N_f)_V \subset SU(N_f)_V$ in the $SU(N_f)$ chiral Lagrangian, using the Cartan embedding to translate between the two formalisms.

Let us consider a U(1) subgroup of $SO(N_f)$ generated by the charge matrix $Q$. We have the following expression for the gauged WZNW term (see Refs. [5, 23]):

$$\tilde{\Gamma} = \Gamma + \int dx^4 x A_\mu J^\mu + \frac{i}{24\pi^2} \int dx^4 \epsilon^{\mu\nu\rho\sigma} (\partial_\mu A_\nu) A_\sigma$$
$$\times \text{Tr}\left[Q^2(\partial_\sigma W)W^\dagger + Q^2 W^\dagger(\partial_\sigma W) + \frac{1}{2}QW^\dagger Q(\partial_\sigma W) - \frac{1}{2}QWQ(\partial_\sigma W^\dagger)\right], \tag{39}$$

where

$$J^\mu = \frac{1}{48\pi^2} \epsilon^{\mu\nu\rho\sigma} \text{Tr}\Big[Q(\partial_\nu W)W^\dagger \cdot (\partial_\rho W)W^\dagger \cdot (\partial_\sigma W)W^\dagger$$
$$+ QW^\dagger(\partial_\nu W) \cdot W^\dagger(\partial_\rho W) \cdot W^\dagger(\partial_\sigma W)\Big]. \tag{40}$$

## 3.3 Anomaly in the ultraviolet

The general expression for the anomalous current in four-dimensional gauge theories can be read off from the triangle graph in Fig. 2,

$$\langle \partial_\mu J^\mu_{\kappa_1} \rangle = \frac{1}{32\pi^2} D_{\kappa_1\kappa_2\kappa_3} \epsilon^{\kappa\nu\lambda\rho} F^{\kappa_2}_{\kappa\nu} F^{\kappa_3}_{\lambda\rho}, \tag{41}$$

where

$$D_{\kappa_1\kappa_2\kappa_3} = \frac{1}{2}\text{Tr}\Big(\{T_{\kappa_1}, T_{\kappa_2}\}T_{\kappa_3}\Big). \tag{42}$$

Let us gauge the U(1) subgroup generated by

$$Q = \begin{pmatrix} 0 & i & 0 \\ -i & 0 & 0 \\ 0 & 0 & 0 \end{pmatrix}. \tag{43}$$

We have $T_{\kappa_2} = T_{\kappa_3} = Q$. The expression for the anomaly coefficient of a generic element of the flavor group $T_{\kappa_3}$ is

$$D_{\kappa_1 QQ} = D_{QQ\kappa_1} = (N_c^2 - 1) \text{Tr}\left[\begin{pmatrix} 1 & 0 & 0 \\ 0 & 1 & 0 \\ 0 & 0 & 0 \end{pmatrix} T_{\kappa_1}\right]. \tag{44}$$

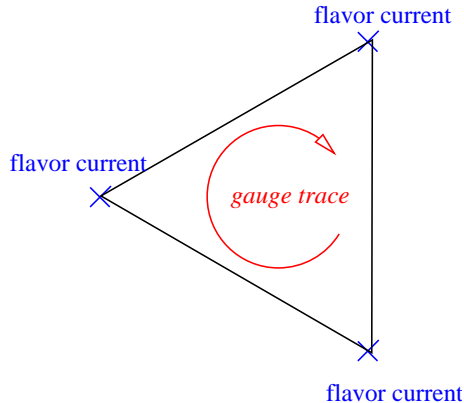

Figure 2: The WZNZ term is responsible for anomaly matching between the ultraviolet (microscopic) theory and the low-energy effective Lagrangian (macroscopic description). The anomalies in question are given by triangle graphs symbolically depicted in this figure, with flavor currents in the vertices – they are blind with respect to the gauge indices. The only information about the gauge structure comes from the trace in the loop.

The fact that $D$ is completely symmetric facilitates the calculation. Let us take
$$T_{\kappa_1} = \begin{pmatrix} 1 & 0 & 0 \\ 0 & 1 & 0 \\ 0 & 0 & -2 \end{pmatrix}, \qquad (45)$$
which corresponds to the Golstone boson $\pi_3$. We then find
$$\langle \partial_\mu J^\mu_{\kappa_1} \rangle = \frac{N_c^2 - 1}{16\pi^2} \epsilon^{\kappa\nu\lambda\rho} F_{\kappa\nu} F_{\lambda\rho}. \qquad (46)$$

At this point we can match this value with the one found from the low-energy theory in Eq. (39). We obtain in this way that the coefficient in front of the WZNW term is
$$k = \frac{N_c^2 - 1}{2}. \qquad (47)$$

The crucial $1/2$ factor comes from the fact that we consider a theory with the Weyl fermions rather than Dirac fermions as is the case in QCD. Note that $k$ is half-integer for $N_c$ even and integer for $N_c$ odd.

# 4 How the WZNW term affects statistics

In order to determine the Skyrmion statistics it is necessary to calculate the value of the WZNW functional $\Gamma$ for a process of a $2\pi$ rotation of the soliton. Let us first briefly review the case of the QCD Skyrmion [6]. We will use the spherical coordinates $(r, \theta, \phi)$ for the spatial directions. In this case the soliton field for $N_f = 2$ can be chosen in the following form:

$$S = \exp\left(i(\vec{n}\,\vec{\tau})\psi(r)\right), \tag{48}$$

where

$$\vec{n} = (\sin\theta\cos\phi,\, \sin\theta\sin\phi,\, \cos\theta),$$

and $\psi(r)$ is a function subject to the following boundary conditions:

$$\psi(r \to 0) = 0, \qquad \psi(r \to \infty) = \pi.$$

Note that in this section the angle $\theta$ has a different meaning compared with other sections.

This solution can be embedded in the effective Lagrangian for $N_f = 3$,

$$Y = \begin{pmatrix} S & 0 \\ 0 & 1 \end{pmatrix}. \tag{49}$$

The explicit expression for $Y$ is

$$Y = \begin{pmatrix} \cos\psi + i\cos\theta\sin\psi & ie^{-i\phi}\sin\psi\sin\theta & 0 \\ ie^{i\phi}\sin\psi\sin\theta & \cos\psi - i\cos\theta\sin\psi & 0 \\ 0 & 0 & 1 \end{pmatrix}. \tag{50}$$

Then the following configuration is considered:

$$B = C^{-1}(t, \rho) \cdot Y(x_i) \cdot C(t, \rho), \tag{51}$$

where

$$C = \begin{pmatrix} 1 & 0 & 0 \\ 0 & \rho e^{it} & \sqrt{1-\rho^2} \\ 0 & -\sqrt{1-\rho^2} & \rho e^{-it} \end{pmatrix}, \tag{52}$$

with the boundaries

$$0 < t < 2\pi, \quad 0 < \rho < 1. \tag{53}$$

It is crucial that at $\rho = 0$ the configuration $B$ is independent of $t$,

$$\begin{pmatrix} \cos\psi + i\cos\theta\sin\psi & 0 & ie^{-i\phi}\sin\psi\sin\theta \\ 0 & 1 & 0 \\ ie^{i\phi}\sin\psi\sin\theta & 0 & \cos\psi - i\cos\theta\sin\psi \end{pmatrix}.$$

For this reason we can think of $(\rho, t)$ as polar coordinates in the plane, where $\rho$ is the radius and $t$ the polar angle.

If we restrict to $\rho = 1$, this field configurations corresponds to a temporal rotation of the Skyrmion from $t = 0$ to $t = 2\pi$ (in Eq. (50) this corresponds to a shift $\phi \to \phi - t$.) If we evaluate the contribution of the WZNW term (36) on the five-dimensional submanifold $B$ with the integration range

$$0 \leq t \leq 2\pi, \quad 0 \leq \rho \leq 1, \quad 0 \leq \psi \leq \pi,$$
$$0 \leq \theta \leq \pi, \quad 0 \leq \phi \leq 2\pi, \tag{54}$$

we arrive at

$$\Gamma = \pi. \tag{55}$$

For odd values of the integer coefficient $k$ of the WZNW term, the contribution to the path integral corresponding to a $2\pi$ rotation of the soliton is $e^{ik\Gamma} = -1$. From the arguments of Ref. [6] we see that this corresponds to a fermionic quantization of the Skyrmion. For even values of $k$, the same contribution is $e^{ik\Gamma} = 1$, which corresponds to bosonic quantization. This concludes our brief review of the situation with the QCD Skyrmions.

In what follows we will carry out the same calculation for the Hopf Skyrmion ($N_f = 2$) embedded in $\mathcal{M}_3$ (see Fig. 3). We will use a field configuration obtained from the one that was exploited in QCD (Eq. (48)) using the Cartan embedding map,

$$S \cdot S^t = \begin{pmatrix} a & b \\ b & a^* \end{pmatrix}, \tag{56}$$

$$\begin{aligned} a &= (\cos\psi + i\cos\theta\sin\psi)^2 - \sin^2\psi\sin^2\theta e^{-2i\phi}, \\ b &= 2i\sin\psi\sin\theta(\cos\psi\cos\phi - \cos\theta\sin\psi\sin\phi). \end{aligned} \tag{57}$$

Let us check that this map, for $N_f = 2$, transforms The QCD Skyrmion into the Hopf Skyrmion. It is convenient to use the coordinates $(\tilde\theta, \tilde\phi)$ defined by

$$\vec{n} = (\sin\tilde\theta\cos\tilde\phi, \cos\tilde\theta, \sin\tilde\theta\sin\tilde\phi). \tag{58}$$

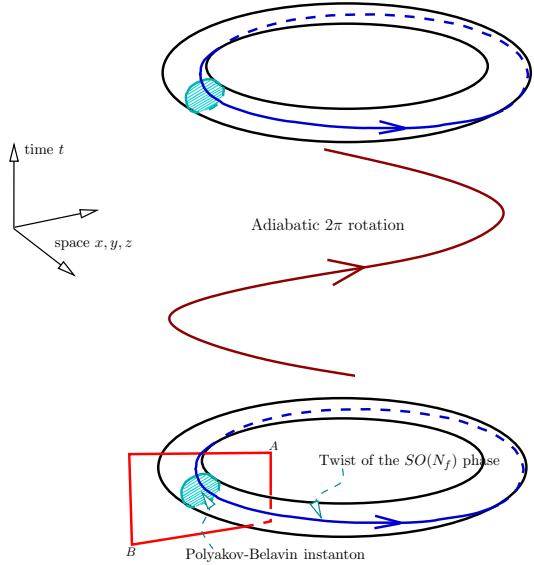

Figure 3: Adiabatic rotation of a Hopf Skyrmion by the $2\pi$ angle. In the explicit calculation presented in this section, the toroidal object is rotated by $2\pi$ along an axis $\hat{x}^3$ which is perpendicular to the symmetry axis $\hat{x}^2$. The contribution of the WZNW term to the action does not depend on the axis that we choose for the $2\pi$ rotation.

The lines
$$\tilde{\theta} = 0, \pi \text{ and } 0 \leq \psi \leq \pi$$
correspond, in the SU(2) Skyrme model, to SO(2) elements. This line is mapped onto identity matrix (which can be identified with the north pole of the sphere $S^2$).

The points
$$\psi = \pi/2, \; \tilde{\theta} = \pi/ \text{ and } 0 \leq \tilde{\phi} \leq \pi$$
corresponding in SU(2) to
$$\begin{pmatrix} i\sin\tilde{\phi} & i\cos\tilde{\phi} \\ i\cos\tilde{\phi} & -i\sin\tilde{\phi} \end{pmatrix},$$
are mapped to minus identity, which can be identified with the south pole of the sphere $S^2$. The linking number of these two links (which are the preimages of the north and the south poles of $S^2$) is one. This shows that the Cartan embedding map is the Hopf fibration with the winding number one. It is easy to check that there is a single twist in the flux tube by considering sections with the fixed $(\psi, \tilde{\theta})$ and $0 < \tilde{\phi} < 2\pi$. This Hopf Skyrmion is a toroidal object with a symmetry axis oriented along the spatial direction $\hat{x}^2$.

One can readily embed this configuration in $\mathcal{M}_3$,

$$Y \cdot Y^t = \begin{pmatrix} a & b & 0 \\ b & a^* & 0 \\ 0 & 0 & 1 \end{pmatrix}. \tag{59}$$

If we calculate $B \cdot B^t$ for $\rho = 1$ we find a shift $\phi \to \phi - t$ in Eq. (59). It corresponds to the adiabatic rotation along the axis $\hat{x}^3$. The toroidal structure in Fig. 3 is rotated along the axis $\hat{x}^3$ perpendicular to the symmetry axis $\hat{x}^2$. For $\rho = 0$ the quantity $B \cdot B^t$ is independent of $t$,

$$\begin{pmatrix} a & 0 & b \\ 0 & 1 & 0 \\ b & 0 & a^* \end{pmatrix}.$$

Therefore, we can still use $(\rho, t)$ as the polar coordinates.

Next, we have to evaluate the WZNW term (36) on $B \cdot B^t$ and perform the integration over all the five variables $(\psi, \theta, \phi, \rho, t)$, with the integration range given in Eq. (54). The result is twice larger than that we get in QCD,

$$\Gamma = 2\pi. \tag{60}$$

With our conventions for the coefficient $k$, it can be integer or half-integer, depending on the number of colors $N_c$. As was discussed in Sect. 3.3, $k$ is half-integer for $N_c$ even and integer for $N_c$ odd. It immediately follows that the Skyrmion is quantized as a fermion for $N_c$ even and as a boson for $N_c$ odd.

## 5 Skyrmions in SO($N_c$) QCD

Now we consider another parental theory: SO($N_c$) gauge theory with $N_f$ Weyl quarks in the vectorial representation. Such a theory can be viewed as a "parental" microscopic theory because it has the chiral symmetry breaking

$$\text{SU}(N_f) \times \mathbb{Z}_{4N_f} \to \text{SO}(N_f) \times \mathbb{Z}_2, \tag{61}$$

which, apart from the discrete factors, is the same as SU($N_c$) Yang–Mills with adjoint Weyl quarks, see Eq. (1).

The low-energy effective Lagrangian is again a nonlinear sigma model with the target space $\mathcal{M}_{N_f}$. The "baryon number" symmetry, which rotates all charge-1 Weyl quarks is also anomalous; the anomaly-free part is $\mathbb{Z}_{4N_f}$. This discrete symmetry is then broken down to $\mathbb{Z}_2$ by the fermion condensate.

There are some differences from adjoint QCD. One of them is that the coupling constant $F_\pi$ scales as $N_c$ rather than $N_c^2$. It means, in turn, that now the Skyrmion soliton is an object whose mass scales as $N_c$. Moreover, the fermion $\psi$ (see Eq. (4)) is absent in the spectrum. Using the arguments discussed in Sect. 3.3, it is straightforward to compute the coefficient $k$ of the WZNW term in the low-energy effective action. The triangle diagram is completely similar to that in the adjoint QCD case. The coefficient comes out different due to a different number of the ultraviolet degrees of freedom. The result is

$$k = \frac{N_c}{2}. \tag{62}$$

From the calculation performed in Sect. 4 it immediately follows that for $N_c$ odd the Skyrmion is a fermion while for $N_c$ even it is a boson.

The Skyrmion in the $\mathrm{SO}(N_c)$ theory had been already matched with the stable particle construction in the microscopic theory. This identification belongs to Witten [6]. He argued that the Skyrmion corresponds to the baryon constructed of $N_c$ quarks,

$$\epsilon_{\alpha_1 \alpha_2 \ldots \alpha_{N_c}} q^{\alpha_1} q^{\alpha_2} \ldots q^{\alpha_{N_c}}. \tag{63}$$

As was discussed in Ref. [24], the gauge theory actually has an $\mathrm{O}(N_c)$ symmetry; the quotient $\mathbb{Z}_2 = \mathrm{O}(N_c)/\mathrm{SO}(N_c)$ acts as a global symmetry group. All particles built with the $\epsilon_{\alpha_1 \alpha_2 \ldots \alpha_{N_c}}$ symbol are odd under this symmetry. This means that the baryon (63) is stable under decay into massless Goldstone bosons while two baryons can freely annihilate. There is a subtlety due to the "Pfaffian" particles (see Ref. [24] for a discussion), since, in addition to (63), we have to consider objects constructed from the $\epsilon_{\alpha_1 \alpha_2 \ldots \alpha_{N_c}}$ symbol, $r$ quarks and $(N_c - r)/2$ gauge field strength tensors,

$$\epsilon_{\alpha_1 \alpha_2 \ldots \alpha_{N_c}} \left( q^{\alpha_1} q^{\alpha_2} \ldots q^{\alpha_r} \right) \left( F^{\alpha_{r+1} \alpha_{r+2}} \ldots F^{\alpha_{N_c-1} \alpha_{N_c}} \right). \tag{64}$$

For $N_c$ odd these objects always contain an odd number $\geq 1$ of quarks; for $N_c$ even they contain an even number of quarks (this number can be zero). It is a dynamical question whether the baryon in Eq. (63) or an object like the one in Eq. (64) is the lowest mass particle. All of these objects have masses that scale as $N_c$ for large $N_c$. They are subject to Fermi statistics for $N_c$ odd and Bose statistics for $N_c$ even. This is consistent with the statistics that we obtained above for the Skyrmion solitons.

From Eqs. (63) and (64) we can infer information about other quantum numbers of the Skyrmion. Its $\mathbb{Z}_2$ fermion number is given by $N_c$ modulo 2 and its flavor representation is contained in the tensor product of $N_c$ vectorial representations. This is consistent with the computation carried out in conventional QCD (with fundamental quarks) in Ref. [25].

By the same token we can argue that there is a similar contribution to the fermion number of the Skyrmion in adjoint QCD, which is proportional to $N_c^2 - 1$. As discussed in Sect. 6, the composite fermion $\psi$ (which is absent in the SO($N_c$) theory) will give an extra contribution to the Skyrmion fermion number shifting it by one unit.

More comments about the difference between the SU($N_c$) and SO($N_c$) theories are in order here. First of all it must be said that our results can be trusted only the the number of colors is sufficiently high. This is because the low-energy effective action has coupling that scales as $\propto 1/\sqrt{N_c}$. The SU($N_c$) theory has a fermionic state $\text{Tr}(F\lambda)$ whose mass scales as $N_c^0$. For this reason it can be included in a low-energy effective Lagrangian and it can have an impact on the Skyrmion quantum numbers. In the SO($N_c$) theory the situation is different. It is possible, although only for odd $N_c$, to construct a gauge invariant fermionic state of the type $\epsilon\psi F \ldots F$. This state will certainly have a mass that grows at least as $N_c^1$. For this reason it cannot be considered in any low-energy effective Lagrangian in the large-$N_c$ limit.

# 6 Skyrmion stability due to anomaly

This section is central for the understanding of the Skyrmion stability in the microscopic theory. In a sense, all previous sections can be viewed as a preparation to this section.

The main outcome of Ref. [12] in the $N_f = 2$ case, is that the Skyrmion, through the Goldstone–Wilczek mechanism [14], acquires a fermion number 1. The reason is as follows. Equation (20) indicates how to express the Hopf charge as a function of the gauge field $A_\mu$. On the other hand, we know that the fermion current has an axial anomaly,

$$\partial^\mu J_\mu^{F0} = \frac{1}{8\pi^2} F_{\mu\nu} \tilde{F}^{\mu\nu} = \frac{1}{4\pi^2} \partial_\mu (\epsilon^{\mu\nu\rho\sigma} A_\nu \partial_\rho A_\sigma), \tag{65}$$

where

$$\widetilde{F}_{\mu\nu} = \frac{1}{2} \epsilon_{\mu\nu\rho\sigma} F^{\rho\sigma}.$$

We have omitted the mass term which explicitly brakes $U(1)_F \to Z_2$.

One must remember that the fermion number is well defined only modulo two. The last equality in Eq. (65) is the well-known expression of the topological density as a total derivative of the Chern–Simons term. Integrating the anomaly equation we arrive at

$$J_\mu^F = J_\mu^{F0} - \frac{1}{4\pi^2} K_\mu, \qquad K_\mu = \epsilon^{\mu\nu\rho\sigma} A_\nu \partial_\rho A_\sigma. \tag{66}$$

This expression is matched with the Hopf number in Eq. (20). We thus find that the fermion $\psi$ transfers one unit of fermion number to the Hopf Skyrmion.

In order to generalize to higher $N_f$ one must consider the triangle anomaly

$$U(1) - SO(N_f) - SO(N_f).$$

The U(1) corresponds to the fermion number. For $SO(N_f)$ we introduce an auxiliary gauge field. The anomaly is

$$\partial_\mu J_\mu^{F0} = \frac{1}{16\pi^2} \text{Tr}(F^{\mu\nu} \widetilde{F}_{\mu\nu}) = \frac{1}{8\pi^2} \partial_\mu K^\mu, \tag{67}$$

where $F_{\mu\nu} = F_{\mu\nu}^k T^k$, with $T^k$ standing for the generators of $SO(N_f)$ (with $\text{Tr}(T_j T_k) = \delta_{ij}$), and $K_\mu$ is given in Eq. (23).

The net effect of the baryon $\psi$ with mass $O(N_c^0)$ is to shift the Skyrmion fermion number by one unit, without changing its statistics. For $N_c$ odd, the Skyrmion is a boson with an odd fermion number. For $N_c$ even, it is a fermion with an even fermion number. The relation between the Skyrmion statistics and fermion number is abnormal. In both cases it is a $\mathbb{Z}_2$-stable object, because in the "perturbative" spectrum the normal relation between the fermion number and statistics takes place.

# 7 Flux tubes

The question of whether or not the flux tubes supported by the chiral Lagrangian are related to confining strings of the underlying microscopic gauge theory was raised by Witten [6] (see also later works [24, 26]). This question has no direct connection to the main issue of the present investigation: the Skyrmion stability in adjoint QCD. However, we would like to add a brief comment which follows from the arguments presented in the previous sections.

First of all, let us note that

$$\pi_2(\mathrm{SU}(N_f)/\mathrm{SO}(N_f)) = \mathbb{Z}_2 \text{ at } N_f \geq 3 \,. \tag{68}$$

This fact implies that the sigma model with the target space (2) does indeed support flux tubes. These flux tubes are $\mathbb{Z}_2$-stable, i.e. a pair of them can annihilate into pions, mesons and "normal" baryons.

The microscopic theory analyzed by Witten was $\mathrm{O}(N_c)$ gauge theory, with quarks in the vector representation of $\mathrm{O}(N_c)$. He gave an indirect argument why the chiral theory flux tube might be a reflection of a confining string of the $\mathrm{O}(N_c)$ gauge theory. His argument is based on the fact that an external probe quark in the spinor representation of $\mathrm{O}(N_c)$ cannot be screened by dynamical quarks in the vector representation. Hence, a confining string is attached to such external probe quark. If we take two strings in the given microscopic theory, they should be attached to two spinorial probe quarks, but two spinors make a tensor which can be screened.

Now we know that one and the same pattern of the chiral symmetry breaking (2) takes place in the $\mathrm{O}(N_c)$ gauge theory with quarks in the vector representation *and* $\mathrm{SU}(N_c)$ gauge theory with quarks in the adjoint representation. However, Witten's argument is totally inapplicable in the latter case. Indeed, in this microscopic theory a probe quark with any number of, say, upper indices $Q^{i_1\cdots i_n}$ and no lower indices cannot be screened by adjoint dynamical quarks. Strings of any $n$-ality, up to $[N_c/2]$, are stable. (Here [...] stands for the integer part.) This tells us that the $\mathbb{Z}_2$-strings supported by the chiral low-energy theory are unrelated to the confinement strings of the corresponding microscopic theories. What phenomenon do they describe? Can the end points of such "pionic" flux tubes be attached to Skyrmions ?

# 8  Conclusions

In this paper we advanced on the way of our studies of the Skyrmion stability in adjoint QCD, extending the results that had been obtained previously to three or more flavors.

The underlying (microscopic) reason for the Skyrmion stability is an odd relationship between the Skyrmion statistics and its fermion number. For $N_c$ odd, the Skyrmion is a boson with an odd fermion number. For $N_c$ even, it is a fermion with an even fermion number. The shift in the fermion number occurs for the same reason as it was first discussed in Ref. [12].

We deal here with the $\mathbb{Z}_2$-stability: a composite state built of two Skyrmions is not stable, in agreement with the fact that

$$\pi_3(\mathcal{M}_{N_f}) = \mathbb{Z}_2, \text{ for } N_f \geq 3.$$

Note that the reason for the Skyrmion stability in another microscopic theory with the same chiral Lagrangian – SO($N_c$) gauge theory with vector quarks – is different. In the latter case the stability is due to $\mathbb{Z}_2 =$ O($N_c$)/SO($N_c$).

Our analysis is valid at large $N_c$. Something peculiar happens when we leave the large-$N_c$ limit and go to small $N_c$. SU($N_c$) adjoint QCD for $N_c = 2$ and the SO($N_c$) gauge theory with vector quarks for $N_c = 3$ are in fact one and the same theory. The SO($N_c$) description in this particular case is better. In this case the fermion $\psi$ coincides with the Pfaffian,

$$\epsilon_{abc} q^a F^{bc},$$

therefore, it does not make sense to introduce it as another independent degree of freedom.

An interesting issue which is not quite solved yet is the interpretation of the Skyrme string associated with (68). Interpretation of the chiral theory strings in adjoint QCD is unclear, and so is the question where these strings end.

# Appendix

In this Appendix we shown that the result of integration of $\Gamma$ on the minimal $S^5$ inside $\mathrm{SU}(N_f)/\mathrm{SO}(N_f)$ (for $N_f > 3$) is $4\pi$.

First let us define what we mean by a "minimal" $S^5$. The relevant homotopy groups are shown in Table 3.

|       | $N_f = 3$ | $N_f = 4$ | $N_f = 5$ | $N_f > 5$ |
|-------|-----------|-----------|-----------|-----------|
| $\pi_5$ | $\mathbb{Z} \otimes \mathbb{Z}_2$ | $\mathbb{Z} \otimes \mathbb{Z}_2 \otimes \mathbb{Z}_2$ | $\mathbb{Z} \otimes \mathbb{Z}_2$ | $\mathbb{Z}$ |

Table 3: The fifth homotopy group for sigma models emerging in Yang–Mills with more than three flavors.

If the WZNW term is integrated over one of the $\mathbb{Z}_2$ factors of the homotopy groups, the result vanishes. Therefore, they are irrelevant to the present discussion. We define a "minimal" $S^5$ as the generator of the $\mathbb{Z}$ factor of each of the $\pi_5$ groups.

Let us consider the exact sequence for the homotopy group. The details are quite complicated. For $k = 3, 5$ we have the following exact sequence:

$$\ldots \to \pi_5\left(\mathrm{SU}(k)\right) \to \pi_5\left(\mathrm{SU}(k)/\mathrm{SO}(k)\right) \to \pi_4\left(\mathrm{SO}(k)\right) \to \pi_4\left(\mathrm{SU}(k)\right) \to \ldots$$

$$\ldots \to \mathbb{Z} \to \mathbb{X} \to \mathbb{Z}_2 \to 0 \to \ldots, \tag{A.1}$$

implying two alternatives,

$$\mathbb{X} = \mathbb{Z} \quad \text{or} \quad \mathbb{X} = \mathbb{Z} \otimes \mathbb{Z}_2\,.$$

In Ref. [18] it is shown that the last option is the correct one. The $k = 4$ case is distinct,

$$\ldots \to \pi_5\left(\mathrm{SU}(4)\right) \to \pi_5\left(\mathrm{SU}(4)/\mathrm{SO}(4)\right) \to \pi_4\left(\mathrm{SO}(4)\right) \to \pi_4\left(\mathrm{SU}(4)\right) \to \ldots$$

$$\ldots \to \mathbb{Z} \to \mathbb{X} \to \mathbb{Z}_2 \otimes \mathbb{Z}_2 \to 0 \to \ldots,$$

which gives us the alternatives

$$\mathbb{X} = \mathbb{Z} \otimes \mathbb{Z}_2 \quad \text{or} \quad \mathbb{X} = \mathbb{Z} \otimes \mathbb{Z}_2 \otimes \mathbb{Z}_2\,.$$

It was shown n Ref. [18] that the last choice is the correct one. For $k = 6$ we get

$$\ldots \to \pi_5\left(\mathrm{SO}(k)\right) \to \pi_5\left(\mathrm{SU}(k)\right) \to \pi_5\left(\mathrm{SU}(k)/\mathrm{SO}(k)\right) \to \pi_4\left(\mathrm{SO}(k)\right) \to \ldots$$

$$\ldots \to \mathbb{Z} \to \mathbb{Z} \to \mathbb{X} \to 0 \to \ldots,$$

which is not enough to find $\mathbb{X}$. In Ref. [18] it was shown that $\mathbb{X} = \mathbb{Z}$. For $k > 6$:

$$\ldots \to \pi_5\left(\mathrm{SO}(k)\right) \to \pi_5\left(\mathrm{SU}(k)\right) \to \pi_5\left(\mathrm{SU}(k)/\mathrm{SO}(k)\right) \to \pi_4\left(\mathrm{SO}(k)\right) \to \ldots$$

$$\ldots \to 0 \to \mathbb{Z} \to \mathbb{X} \to 0 \to \ldots,$$

which gives $\mathbb{X} = \mathbb{Z}$.

A common feature of all these exact sequences is that the generator of $\pi_5\left(\mathrm{SU}(k)\right)$ is mapped onto the generator of the $\mathbb{Z}$ factor of $\pi_5\left(\mathrm{SU}(k)/\mathrm{SO}(k)\right)$. A concrete realization of this map is given by the Cartan embedding, interpreted as a map from $\mathrm{SU}(k)$ to the space of symmetric Hermitian matrices, which is a realization of $\mathrm{SU}(k)/\mathrm{SO}(k)$,

$$U \to W = U \cdot U^t.$$

Thus, in order to calculate $\Gamma$ on the minimal $S^5$ inside the space of the symmetric Hermitian matrices, we need to calculate $\Gamma$ on the Cartan embedding image of a generator of $\pi_5\left(SU(k)\right)$. It is sufficient to carry out the explicit calculation for $k = 3$ (for $k > 3$ the result is given by a trivial embedding of the cycles used for $k = 3$).

If we parameterize $S^5$ as a sphere in $\mathbb{C}^3$ with the coordinates $(z_1, z_2, z_3)$, the generator of $\pi_5\left(SU(3)\right)$ is (see Ref. [18])

$$\eta(z) = z\, z^t + \begin{pmatrix} 0 & -\bar{z}_3 & \bar{z}_2 \\ \bar{z}_3 & 0 & -\bar{z}_1 \\ -\bar{z}_2 & \bar{z}_1 & 0 \end{pmatrix}. \tag{A.2}$$

It is then straightforward to calculate the integral of the WZNW term on this $S^5$ cycle and on the Cartan embedding image. Namely,

$$\Gamma_1 = \int_{S^5} S_{\mathrm{WZNW}}\left(\eta(z)\right) = 2\pi, \quad \Gamma_2 = \int_{S^5} S_{\mathrm{WZNW}}\left(\eta(z) \cdot \eta(z)^t\right) = 4\pi. \tag{A.3}$$

This proves that the minimal $S^5$ in the subspace of the symmetric Hermitian generators is twice the minimal $S^5$ in $\mathrm{SU}(N_f)$.